\documentclass[%
 reprint,
superscriptaddress,
 amsmath,amssymb,
 aps,
]{revtex4-1}

\usepackage{graphicx}
\usepackage{dcolumn}
\usepackage{bm}
\usepackage{enumitem}
\usepackage{color,xcolor}

\begin{document}

\preprint{APS/123-QED}

\title{Ptychography Intensity Interferometry Imaging for Dynamic Distant Object}

\author{Yuchen He$^\dag$}
\author{Yuan Yuan$^\dag$}
\author{Hui Chen $^\dag *$}
\email{chenhui@xjtu.edu.cn. \\ $\dag$ These three authors contributed equally.}
\author{Huaibin Zheng}
\author{Jianbin Liu}
\author{Zhuo Xu}
\affiliation{%
Electronic Materials Research Laboratory, Key Laboratory of the Ministry of Education and International Center for Dielectric Research, School of Electronic Science and Engineering, Xi'an Jiaotong University, Xi'an 710049, China\\
}%

\date{\today}

\begin{abstract}
    As a promising lensless imaging method for distance objects, intensity interferometry imaging (III) had been suffering from the unreliable phase retrieval process, hindering the development of III for decades. Recently, the introduction of the ptychographic detection in III overcame this challenge, and a method called ptychographic III (PIII) was proposed. We here experimentally demonstrate that PIII can image a dynamic distance object. A reasonable image for the moving object can be retrieved with only two speckle patterns for each probe, and only 10 to 20 iterations are needed. Meanwhile, PIII exhibits robust to the inaccurate information of the probe. Furthermore, PIII successfully recovers the image through a fog obfuscating the imaging light path, under which a conventional camera relying on lenses fails to provide a recognizable image.

\end{abstract}

\maketitle


\section{\label{sec:level1}INTRODUCTION}
Most conventional imaging methods for distance objects are based on the first-order interference, such as amplitude (phase) interferometry that has been widely used in astronomy for a century\cite{Michelson1921Measurement,Pease1930The,Pease1931Interferometer,Labeyrie1975Interference}. However, Amplitude interferometry (AI) is vulnerable to atmospheric turbulence, and the transferring electric fields in long distance is complex and of high cost\cite{Goodman1976Comparative,Currie1974Four,Breckinridge1994Amplitude,Iaconis1998Spectral,Henderson1998Manual}. These drawbacks limit its applications for very long baselines at optical wavelengths\cite{Breckinridge1990Amplitude,Greenaway1978On,Lebohec2010Stellar,Dravins2015Long}. In 1956, intensity interferometry (II) was invented, which exploits the second-order correlations of the intensity fluctuations in two distance locations\cite{Brown1956A,Hanbury1994Correlation,Brown1975The}. II electronically (rather than optically) connects independent telescopes, which circumvents the drawbacks in AI\cite{Twiss1969Applications}. A typical II system consists of a set of detectors which independently record the time-varying intensities at different locations. These detected intensities are then used to calculate the correlations between them, yielding the power spectrum of an distant object. Since it avoids transmitting light fields from one distant location to another, it not only simplifies the detection process and reduce the cost\cite{Baym1998The}, but also is insensitive to either atmospheric turbulence or telescopic optical imperfections. Thus, these great advantages enable very long baselines (kilometers) observations in short optical wavelengths with a microarcsecond ($\mu as$) resolution\cite{Dravins2010Stellar,Lawson2000Principles,Mccarthy1975Initial,Staelin1978Long}.

Between 1963 to 1974, II had been applied to measure the angular diameters of 32 stars and achieved a resolution as high as 0.4 milliarcsecond\cite{Brown1956A,Hanbury1994Correlation,Brown1975The}. However, this technique was then dormant, due to its low signal-to-noise ratio (SNR) and long measurement time\cite{Labeyrie2006An}. Recently, detection technology has been greatly developed, which will promisingly increase SNR and shorten the measurement time with reasonable cost. Therefore, II has been re-attracted a lot of attentions recently. Especially, an international project called Cherenkov Telescope Array has been being attempting to exploit air Cherenkov telescopes to construct kilometers baselines of II, providing resolution approaching $30\mu as$\cite{Malvimat2014Intensity}.

On the other hand, the measurement of II only yields the spatial Fourier magnitudes of an object's intensity distribution\cite{Brown1975The}. After combined with phase retrieval processes\cite{Fienup1978Reconstruction,Fienup1982Phase,Wackerman1986Phase,holmes_investigation_2004,rogers_two-dimensional_2010}, II is able to reconstruct the image of an distance object. However, the mostly employed algorithms, such as Error Reduction (ER) and Hybrid Input-Output (HIO), are struggling with unreliable and time-consuming problems, making recovering a complex object impractical\cite{Fienup1978Reconstruction,Fienup1982Phase,Wackerman1986Phase}. Recently, ptychographical scanning was introduced into II which can circumvent this challenging\cite{Chen}, and a new imaging method called ptychography intensity interferometry imaging(PIII) was proposed to image complex-shaped objects\cite{Chen}.
Moreover, due to exploiting the second-order correlation, PIII has several advantages in comparison to ptychographical iterative engine (PIE) in coherent diffraction imaging. PIE requires the precise information of the probe function\cite{Huang2009,Guizar2008,Shenfield2011,Maiden2012,Beckes2013,Zhang2013} as well as the accurate distance between an object and a detector\cite{Maiden2012,Bian2013,Dou2017,Loetgering2018}. In contrast, PIII neither needs the information of the detection distance, nor the precise knowledge of the probe function. Instead, it just uses the approximate region of the probe. These advantages make PIII suitable for imaging distant object. In this article, we demonstrate imaging of a dynamic object $20\; meters$ away from an imager built based on PIII. Different than previously proposed PIII, we design a new algorithm aim to imaging a dynamic object with less samples of illumination speckles onto the object. Moreover, we experimentally prove that PIII has an ability of imaging through a thick fog under which the traditional lens imaging system could not provide a feasible image. This work proves that PIII can work well with dynamic distant objects, which dramatically lows down the requirement for the knowledge of the probe and the experimental setup, resulting in its ability of imaging through turbulence.

\section{Experiments and Methods}
\subsection{The Imager}
\begin{figure}[h]
\includegraphics[width=8.5 cm]{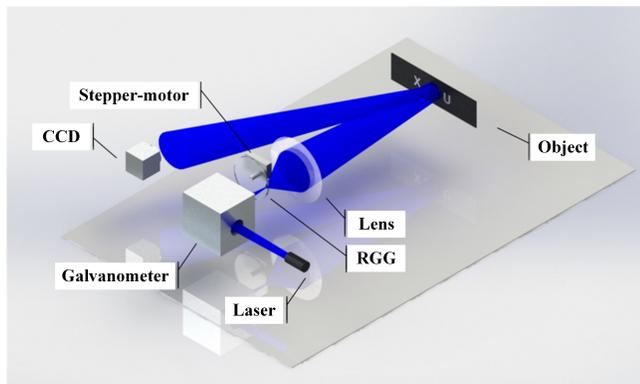}
\caption{\label{light-path} The schematic diagram of the light path.}
\end{figure}

The schematic of the experiment is shown in Fig.~\ref{light-path}. The object is made of four letters ``2GNS''. Each letter is of $\sim7\times7\;cm$, which is placed 20 meters away from the imager. The imager consists of illumination and detection parts. The detection part is simply a bare CCD. In the illumination part, a laser beam is deflected by a galvanometer onto a rotation ground glass (RGG), which is then collimated by a telescope system. This system generates a sequence of light speckle patterns within a circular area on the object plane while the ground glass is rotating. The circular area, acting as a probe, is controlled to $10\;cm$ in diameter.  While the object (mounted on a stage) is making a translational motion, it is illuminated part by part, realizing a ptychographical scan. The rotating speed of RGG is much faster than the motion of the object. So, in each scan, the CCD can capture multiple speckle patterns constructed by the reflected field from the object.

\begin{figure}[h]
    \includegraphics[width=8.5 cm]{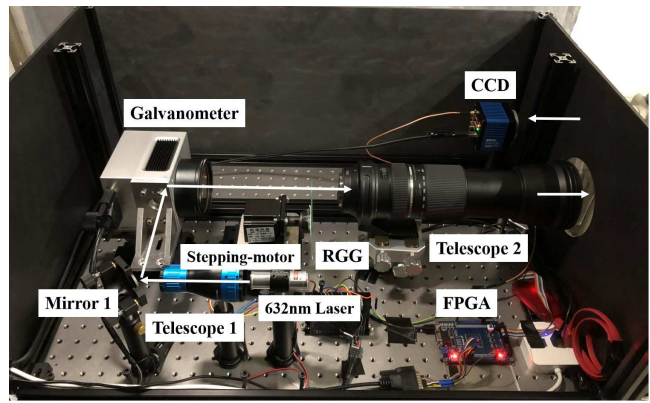}
    \caption{\label{fig:imager} The physical map of the imager. Telescope 1 expands the laser beam to a $1\;cm$ diameter one. Telescope 2 projects the scattering light from RGG to scene. Meanwhile, it controls the illumination area on the object plane. The wavelength of the laser beam is 632 nm.}
    \end{figure}

On the other hand, if the object is static, the galvanometer can deflect the laser beam two-dimensionally, making a ptychographical scan on the object. The whole imager is controlled by a field programmable gate array (FPGA) circuit to synchronize the illumination and detection. Fig.~\ref{fig:imager} shows the internal structure of the imager.

\subsection{Imaging Methods}
Let's assume the illumination filed on the object plane is $M({\mathbf \xi})$. The ensemble average of its squared modulus is the profile of the illumination area, i.e., $P({\mathbf \xi})=\langle |M({\mathbf \xi})|^2\rangle$, defining a probe function for the ptychographical scan.  The modulation function for the field incident on the object is assumed as  $\Gamma({\mathbf \xi}-{\mathbf v}t)$, where ${\mathbf v}$ is the velocity of the object. The reflectance of the object is  $O({\mathbf \xi}-{\mathbf v}t)=|\Gamma({\mathbf \xi}-{\mathbf v}t)|^2$. Therefore, the reflected field from the object that eventually arrives at the CCD can be written as,
\begin{equation}
    E({\mathbf r},t)=\int M({\mathbf \xi})\cdot \Gamma({\mathbf \xi}-{\mathbf v}t)\cdot g({\mathbf r}-{\mathbf \xi};z)d{\mathbf \xi},
\end{equation}
where $g({\mathbf r}-{\mathbf \xi})$ is a green function that propagates the field from the object plane to the detection plane at a distance $z$.

In the experiment, the object's translational speed is less than $10\;mm/s$. The exposure time of the CCD is $2\;ms$. In a time of $\Delta t\le 20\;ms$, the CCD can capture 20 frames of different speckle patterns caused by RGG. Within such a short time, the object is effectively static and the speckle patterns can be denoted as ${I_j({\mathbf r},t)=|E_j({\mathbf r},t)|^2}$. The average autocorrelation of the speckle patterns is calculated as
\begin{equation}\label{eq:G2}
    G^{(2)}(\Delta{\mathbf r},t)=\sum_{j=1}^N{[I_j\star I_j](\Delta{\mathbf r},t)}.
\end{equation}
Theoretically, it is equivalent to the second-order correlation, which can be calculated using the Van Cittert-Zernike theorem \cite{Cittert1934Die,Zernike1938The},
\begin{align}\label{eq:VCZ}
    G^{(2)}(\Delta{\mathbf r},t)&=\langle I({\mathbf r},t)I({\mathbf r}+\Delta{\mathbf r},t) \rangle/|\langle I({\mathbf r},t)\rangle|^2\cr
    &=1+|\gamma(\Delta{\mathbf r},t)|^2,\cr
    \gamma(\Delta{\mathbf r},t)&=\int \left[P({\mathbf \xi})\cdot O({\mathbf \xi}-{\mathbf v}t)\right]\cdot exp\{ik\mathbf \xi\cdot \Delta{\mathbf r}\}.
\end{align}
This indicates that the intensity correlation reveals the power spectrum of the part of the object within the probe at time $t$.

While the object is moving, the second-order correlations at different times yield the power spectrums of different parts of the object. In the experiment, we select
$N$ parts that covered by the probe at $N$ different times, $\{O({\mathbf \xi}-{\mathbf v}t_m)\}$ with $m=1,2,...,N$. The corresponding power spectrums $\{|\gamma(\Delta{\mathbf r},t_m)|^2\}$ are calculated with Eq.(\ref{eq:G2}), which are applied to the following reconstruction algorithm that designed specifically for dynamic objects.
\begin{enumerate}[itemindent=0em]
    \renewcommand{\labelenumi}{\theenumi)}
    \setlength{\hangindent}{0em}
    \item start from the first big loop of $r=1$ and the first part of $m=1$ with a guess for a whole object function $O_r({\mathbf \xi})\}$.
    \item calculate the Fourier transform $F_{m,r}=\mathcal{F}\{P({\mathbf \xi})\cdot O_r({\mathbf \xi}-{\mathbf v}t_m)\}$.
    \item reconstruct the Fourier transform with the corresponding measured power spectrum:
    $$F'_{m,r}=|\gamma(\Delta{\mathbf r},t_m)|e^{ik\cdot arg\{F_{m,r}\}},$$
    where $arg\{F\}$ is an operator to take the phase of $F$.
    \item compute the inverse Fourier transform : $\Theta_{m,r}({\mathbf \xi}   )=\mathcal{F}^{-1}\{F'_{m,r}\}$.
    \item apply realness and non-negativity constrains to $\Theta_{m,r}({\mathbf \xi})$, yielding $\Theta'_{m,r}({\mathbf \xi})$ :
    $$ \Theta'_{m,r}({\mathbf \xi})=\left\{
    \begin{aligned}
    &Re\{\Theta_{m,r}({\mathbf \xi})\} &,\;P({\mathbf \xi})=1\cap [Re\{\Theta_{k,j}({\mathbf \xi})\} \geqslant\ 0] \\
    &0  &,\;P({\mathbf \xi})=0 \cup [Re\{\Theta_{k,j}({\mathbf \xi})\}<0 ]
    \end{aligned}
    \right.
    $$
    where $Re\{\}$ is to compute the real parts.
    \item update $O_r({\mathbf \xi})$ with $\Theta'_{m,r}({\mathbf \xi})$:
    $$ O'_r({\mathbf \xi}) = O_r({\mathbf \xi})\cdot \left[1-W({\mathbf \xi}+{\mathbf v}t_m)\right] + \Theta'_{k,j}({\mathbf \xi}+{\mathbf v}t_m)\cdot W({\mathbf \xi}+{\mathbf v}t_m). $$
    Here $W({\mathbf \xi})$ is 2D Tukey window with a window length equal to the size of the probe, which is used to smooth the fusion of the recovered parts.
    \item if $m$ is the last part, goto 8); otherwise goto 2) to compute the next part.
    \item after all the parts has been gone through, start the next big iteration ($r\rightarrow r+1$) and going to 2).
    \item if a satisfactory image is obtained or maximum iteration number is reached, exit the calculation.
  \end{enumerate}

It is worthwhile to note that Eq. (\ref{eq:VCZ})  and the whole reconstruction process do not contain the distance between the object and the detector. Based on the Van Cittert-Zernike theorem\cite{Goodman2000}, when the scattering satisfies the ergodic-like condition\cite{Freund1990Looking}, the second-order correlation yields the power spectrum of the object's intensity function, in spite of it is located in the near- or far-field region. Therefore, in principal, PIII does not need to know the detection distance at all, which makes PIII very practical especially for imaging distant objects where the distance is usually hard to be precisely determined.

\section{Experimental Results}
\subsection{The autocorrelation}
\begin{figure}[h]
\includegraphics[width=8.5 cm]{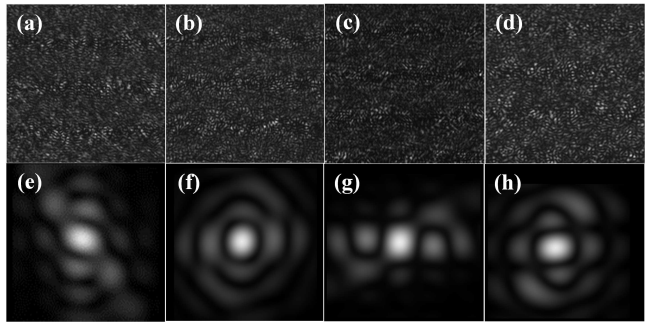}
\caption{\label{ac-results} The autocorrelation results, each calculated from 500 realizations of the speckles. (a)-(d). A speckle pattern of letter '2', 'G', 'N' and 'S', respectively. (e)-(h). The calculated autocorrelation of letter '2', 'G', 'N' and 'S', respectively. }
\end{figure}

In each probe (the object was effectively static during this moment), we took a sequence of speckle patterns and calculated the average autocorrelation using Eq. (\ref{eq:G2}).  To investigate the autocorrelation experimentally, we slowed down the speed of the object, so that we were able to capture 500 speckle patterns for each probe. Fig.~\ref{ac-results}.(a)-(d) show the speckles pattern captured by the CCD. Fig.~\ref{ac-results}.(e)-(h) represent the autocorrelation results. It can be seen that the object's image can not be seen from the speckle pattern, but its autocorrelation reveals the power spectrum of the object.

\subsection{Image Reconstruction for the moving object }
The autocorrelation with 500 speckle patterns provides a clear power spectrum of the object. Thus, the image can be recovered with the traditional phase retrieval algorithm, such as the HIO or the ER. However, it is no realistic to take 500 patterns for each probe, since it requires the object moving in a very slow speed. For example, the exposure time of the CCD is $2\;ms$. It takes $1\;s$ for 500 patterns. To guarantee the object is effective static during the measurement, it should move less than a one-thousandth width of the probe which is $\sim 100\mu m$, i.e., a moving speed less than $100\mu m/s$. If we dramatically reduced the number of the captured patterns for a probe, for instance less than five patterns (the moving speed can be up to $10mm/s$), the calculated autocorrelation would be noisy and it would be very hard to use the HIO (or ER) to reconstruct a faithful result.

\begin{figure}[h]
    \includegraphics[width=7 cm]{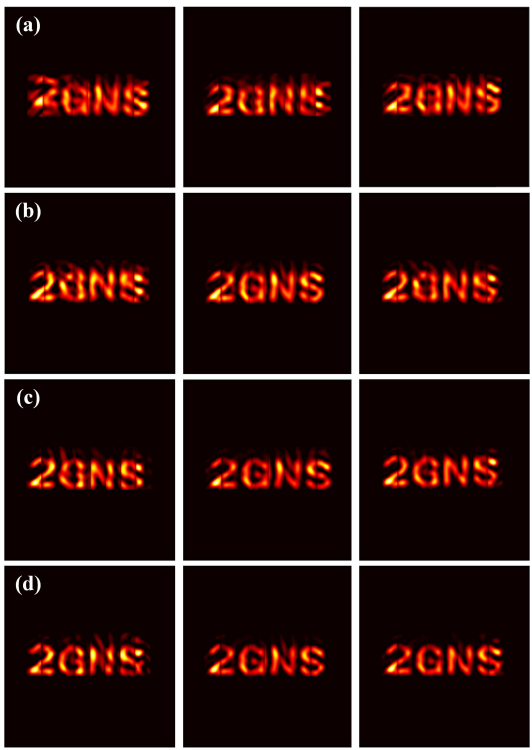}
    \caption{\label{fig:results} Moving objects reconstruction results. (a) Images of  ``2GNS'' reconstructed from the data with 2 speckles captured in each probe. (b) Reconstructed images with 100 speckles captured in each probe. (c) Reconstructed images with 100 speckles captured with 10, 20 and 50 iterations. (d) Reconstructed images with 500 speckles captured in each probe. In each row, from left to right are the results with 10, 20 and 50 iterations, respectively.}
    \end{figure}

\begin{figure}[h]
\includegraphics[width=9 cm]{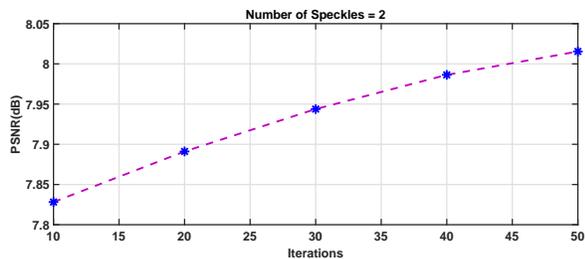}
\caption{\label{psnr} The PSNR results of two speckles.}
\end{figure}

In the experiment, we used just two patterns for each probe, and found that our ptychographic reconstruction algorithm proposed in the above section was able to recover a correct image, as shown in  Fig.~\ref{fig:results}(a).  We also implemented the experiment with 10, 100, 500 patterns for each probe for comparison, as shown in Fig.~\ref{fig:results}(b)-(d). It can be seen that, the reconstruction with two patterns after 10 iterations can still yield a correct result, although the image quality is degraded in comparison with the cases with more patterns for each probe. Nevertheless, the increase of the iteration number can improve the quality. Moreover, we gave a quantitative analysis. We studied the change of peak-signal-to-noise-ratio (PSNR) with the number of iterations in the case of 2 speckles. Fig.~\ref{psnr} shows that as the number of iteration increases, PSNR also increases.

It is worthwhile to point out that, we used a 2D Tukey window as a probe function in the reconstruction algorithm. The window size is a little larger than the actual size of the probe. This window is different than the actual probe: it is roughly rectangular in shape, while the probe is circular. With such a quite inaccurate probe function, PIII still recovers clear images, indicating its robustness to inaccurate information of the probe.

\subsection{Imaging through turbulence}
\begin{figure}[h]
\includegraphics[width=8.5 cm]{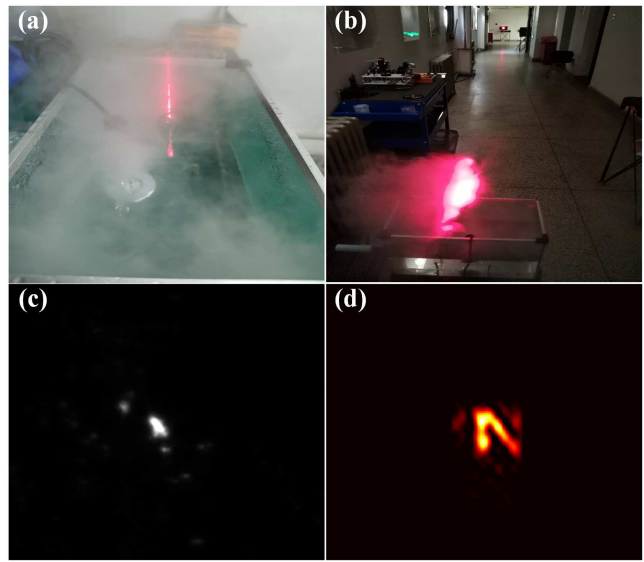}
\caption{\label{anti} The experiment of imaging through turbulence. (a) Water mist effect. (b) The overlook of the experiment with thick fog in between the imager and the object. . (c) Direct image with a traditional camera. (d) The reconstructed image with the PIII method.}
\end{figure}

Furthermore, we investigate the imaging in turbulence condition using the PIII method. Fig.~\ref{anti} shows experimental diagram and results. We used the ultrasonic atomizer to generate fog between the imager and the object. Because the fog scattered the light passing through, using a traditional camera is hard to capture a feasible image of the object, as shown in Fig.~\ref{anti}(c). However, the PIII method can recover a reasonable image. Fig.~\ref{anti}(d) shows the reconstruction result for imaging throught the fog.

\section{Conclusions}
In this paper, we demonstrated a PIII experiment for dynamic distant object 20 meters away. We designed an imager that can implement ptychographic II measurements for a distance object. While the object was translational moving, the imager captured a sequence of speckle patterns. With the proposed PIII algorithm designed for dynamic object, the object was recovered. The number of the speckle patterns for each probe can be low to 2, with which the PIII algorithm can still yield a reasonable image in 10 iterations. Furthermore, we investigated its capability of  imaging through turbulence. With fog obfuscating the imaging light path, a traditional camera relying on lenses failed to provide a recognizable image. In contrast, the PIII method reconstructed a reasonable image through the fog. This research paves a way to the practical application of intensity interferometer imaging.

\begin{acknowledgments}
This work is supported by the National Natural Science Foundation of China (Grant No. 61901353 \& 11503020), 111 Project of China (Grant No. B14040), Key Innovation Team of Shaanxi Province (Grant No. 2018TD-024) and the Fundamental Research Funds for the Central Universities (Grant No. xjh012019029).
\end{acknowledgments}

\nocite{*}

\bibliographystyle{apsrev4-2}

\bibliography{PIII}

\end{document}